\documentclass{aa} 

\newcommand{\ergs}{\mbox{erg s$^{-1}$}}
\newcommand\lsim{\lower0.5ex\hbox{$\; \buildrel < \over \sim \;$}}

\usepackage{graphicx}
\authorrunning{Liu \& Li}
\titlerunning{Faint X-ray sources in GCR}

\begin{document}
   \title{A population synthesis study on the faint X-ray
   sources in the Galactic center region}

   \author{X.-W. Liu and  X.-D. Li
          }

   \offprints{X.-W. Liu}

   \institute{Department of Astronomy, Nanjing University,
               Nanjing 210093, China\\
              \email{liuxw@nju.edu.cn, lixd@nju.edu.cn}
                        }

   \date{Received March 29, 2005; accepted November 10, 2005}


  \abstract
   {Recent \emph{Chandra} observations of the Galactic
center region (GCR) have uncovered a population of faint discrete
X-ray sources.  A few theoretical works have been made to
investigate the nature of these sources.}
   {We examine the contributions and luminosity functions of various
kinds of candidate objects which are proposed either by previous
authors or by ourselves.}
   {We conduct a population synthesis calculation based on Hurley et
al.'s rapid binary evolution code.  Several candidate models, i.e.
wind-accreting neutron stars, intermediate polars, low mass X-ray
binaries, young pulsars and massive stars with strong winds, are
incorporated into our calculation.  We also take the geometric
effect of the accretion disk into account for Roche lobe overflow
X-ray binaries.}
   {Our results show that neutron star low-mass X-ray binaries contribute
   significantly to the observed sources.  We also point out that wind-accreting neutron
   stars contribute negligibly to these sources due to propeller effect,
   and the intermediate polars play a minor role in accounting for the faint
   X-ray sources in both Wang et al. and Muno et al. survey.
  It should be mentioned that the majority of the sources
   in the survey field of Wang et al. are
   still beyond our expectation.}
   {}


   \maketitle
%

\section{Introduction}

Instrumental development always leads people to study sources
looking fainter. Recent {\em Chandra} observations of the Galactic
center region (GCR) have uncovered a population of faint discrete
X-ray sources.  Wang, Gotthelf, \& Lang (\cite{Wang}) have
reported hundreds of X-ray sources with luminosities $L_{\rm
X}\sim 10^{33}- 10^{35}\,\ergs$ in the GCR which contains $\lsim
1\%$ of the total Galactic population (Pfahl et al. \cite{Pfahl},
hereafter PRP02). Many of these sources are emitting in a hard
energy ($>3$ keV) band since the X-ray absorption, varying across
the field, is particularly high in the $1-3$ keV band. Muno et al.
(\cite{Muno03}) presented a catalog of 2357 point sources detected
during their 590 ks {\em Chandra} observations of the $17'\times
17'$ field around Sgr A$^{\ast}$. It was estimated that there are
$\sim 2000$ individual point-like sources with luminosities
$L_{\rm X}\sim 10^{30}-10^{33}\,\ergs$ lying within a cylinder of
radius 20 pc and  depth 440 pc centered on the Galactic center at
a distance of 8.5\,kpc (Muno et al. \cite{Muno04a}). The spectra
of these sources can be fitted by an absorbed power law with
photon index $\Gamma$. More than 1000 sources have relatively hard
spectra with $\Gamma<1$.

The nature of the faint X-ray sources in the GCR has been
investigated by several authors. Pfahl et al. (\cite{Pfahl})
attributed a significant fraction of the X-ray sources revealed by
Wang et al. (\cite{Wang}) to wind-accreting neutron stars (WNSs).
By calculating the flux distribution of WNSs, they concluded that
the detected number ranges from ten to several hundred. The
similar idea for pre-low-mass X-ray binaries was suggested by
Willems \& Kolb (\cite{Willems}, see however, Popov \cite{Popov}).
Muno et al. (\cite{Muno04a}) proposed that magnetized cataclysmic
variables (CVs), referred to as polars and intermediate polars
(IPs), could account for $\sim 1000-2000$ X-ray sources in their
survey. The population synthesis work by Belczynski \& Taam
(\cite{Belczynski}, hereafter BT04) suggested that transient
neutron star low-mass X-ray binaries (NS LMXBs) may contribute
primarily to the X-ray source population in Muno et al. survey.

In this paper, we employed the evolutionary population synthesis
(EPS) method to calculate the expected numbers  and luminosity
distributions of various types of candidates that shine X-rays in
Wang et al. and Muno et al. surveys. We examined the spin
evolution of WNSs proposed by PRP02, including the ejector,
propeller and accretor stages (Davies \& Pringle \cite{Davies},
hereafter DP81; Lipunov \cite{Lipunov}; Ikhsanov
\cite{Ikhsanov01}, \cite{Ikhsanov02}). Besides traditional X-ray
binaries and CVs, we also considered the contribution from
rotation-powered pulsars and massive stars with strong winds. We
describe the population synthesis method and the input physics for
various types of X-ray sources in our model in \S 2. The
calculated results are presented in \S 3. Our discussions and
conclusions are in \S 4.

\section{Model description}

We have used the EPS code developed by Hurley et al.
(\cite{Hurley00}, \cite{Hurley02}) to calculate the expected
numbers in the Galactic disk for various types of single and
binary X-ray source populations. This code incorporates evolution
of single stars with binary-star interactions, such as mass
transfer, mass accretion, common-envelope (CE) evolution,
collisions, supernova kicks, tidal friction and angular momentum
loss mechanics. Most of our adopted parameters are the same as
those described in Hurley et al. (\cite{Hurley02}). We assume all
stars are born in binary systems. The initial mass function (IMF)
of Kroupa, Tout, \& Gilmore (\cite{Kroupa}) is taken for the
primary's mass ($M_1$) distribution (discussed below in \S 2.3).
For the secondary stars ($M_2$), we assume a uniform distribution
of the mass ratio $M_2/M_1$ between 0 and 1. A uniform
distribution of $\ln a$ is also taken for the binary separation
$a$. The star formation rate parameter is $S=7.6085\,{\rm
yr}^{-1}$, corresponding to a rate of $\sim 0.02\,{\rm yr}^{-1}$
for core-collapse supernovae (SNe) in our Galaxy, assume all the
stars with masses $>8M_{\odot}$ die through SNe. During the SN
explosions, a kick velocity $v_{\rm k}$ is imparted on the newborn
compact stars with the Maxwellian distribution
\begin{equation}
   P(v_{\rm k})=\sqrt{\frac{2}{\pi}}\frac{v^{2}_{\rm k}}{\sigma^{3}}
   exp(-\frac{v^{2}_{\rm k}}{2\sigma^{2}})
\end{equation}
where $\sigma=265\,{\rm kms}^{-1}$(Hobbs et al. \cite{Hobbs}) or
$190\,{\rm kms}^{-1}$(Hansen \& Phinney \cite{Hansen}). The CE
parameter $\alpha_{CE}$ is set to be 1, if not otherwise
mentioned. The velocity $v_{W}$ of winds from massive stars is
taken to be $v_{W}=\sqrt{2\beta (GM/R)}$, where the value of
$\beta$ depends on the spectral type of the stars. Hurley et al.
(\cite{Hurley02}) suggested it to be in the range of
0.125\,-\,7.0.

\subsection{Physical Models for X-ray Sources}

\subsubsection{WNSs}

Similar as in PRP02, in our model a WNS is a neutron star (NS)
accreting wind material from an intermediate- or high-mass ($>3
M_{\odot}$) un-evolved companion. We assume that the wind is
steady and spherically symmetric. The standard Bondi-Hoyle
(\cite{Bondi}) accretion formula is employed to calculate the mass
transfer rate to the NS.  Be/NS binaries are not considered in the
present model, due to their high eccentricities and the unclear
structure of the Be star's equatorial winds.  PRP02 simply assumed
that a WNS always accretes from the wind of its binary companion,
ignoring its previous evolution as an ejector and propeller.
Obviously the expected number of WNSs would be less than in PRP02,
depending on the duration of the latter stages. The following
equations give the estimates of the timescales for the ejector
(a), supersonic (c) and subsonic (d) propeller phases for a natal
NS in a wind environment (DP81; Ikhsanov \cite{Ikhsanov02}),
   \begin{equation}
      \tau_{a}\simeq4.8\times10^{6}\mu^{-1}_{31}\dot{M}^{-1/2}_{17}
      I_{45}V^{-1}_{7}\,\mbox{yr},
   \end{equation}
   \begin{equation}
      \tau_{c}\simeq10^{6}\mu^{-1}_{31}\dot{M}^{-1/2}_{17}I_{45}
      V^{-1}_{7}\,\mbox{yr},
   \end{equation}
   \begin{equation}
      \tau_{d}\simeq10^{3}\mu^{-2}_{31}mI_{45}P_{100}\,\mbox{yr},
   \end{equation}
where $\mu=10^{31}\mu_{31}$ Gcm$^3$ is the magnetic moment,
$P=100P_{100}$ s the spin period, $M_{\rm NS}=mM_{\odot}$ the
mass, $I=10^{45}I_{45}$ gcm$^2$ the moment of inertia of the NS,
respectively. $V=10^7V_7$ cms$^{-1}$ is the relative velocity
between the NS and the wind.  The mass capture rate of the NS
$\dot{M}=10^{17}\dot{M}_{17}$ gs$^{-1}$ is given by $\dot{M}=\pi
R^{2}_{\rm a}\rho_{\infty}V$, where $R_{\rm a}=2GM_{\rm NS}/V^{2}$
is the accretion radius and $\rho_{\infty}$ is the density of the
surrounding gas.  When the spin period increases to the brake
period $P_{\rm
br}=100\mu^{16/21}_{31}\dot{M}^{-5/7}_{17}m^{-4/21}\mbox{s}$, the
NS switches from subsonic propeller to accretor.

In the ejector stage, the transferred wind material is outside the
light cylinder. The X-ray emission is generated from the
rotational power of the NS. This will be discussed in \S 2.1.4.
When the pressure of the ejected particles from the neutron star
can no longer balance the ram pressure of the wind gas, the gas
will penetrate into the light cylinder and interact with the
star's magnetosphere, giving rise to X-ray emission. This
propeller stage can be further divided into supersonic and
subsonic stages (DP81). The corresponding X-ray luminosities are
evaluated according to the work by Ikhsanov (\cite{Ikhsanov01}).

Finally when the magnetic pressure is less than the ram pressure
of the falling material and the material at the magnetospheric
boundary is cool enough so that the interchange instabilities can
be triggered, the NS enters the accretor stage, and all the
material captured by the NS is assumed to reach its surface.  We
then use $L_{\rm X}=\eta\dot{M}c^2$ with $\eta\simeq 0.1$ to
calculate the X-ray luminosity.

The parameters of the natal NSs are set as follows. The initial
spin periods $P$ and magnetic fields $B$ are chosen so that $\log
P$ and $\log B$ are distributed normally with a mean of $-2.3$ and
$12.5$ respectively, and a standard deviation of 0.3.

\subsubsection{ Magnetic CVs}

Muno et al. (\cite{Muno04a}) suggested that the majority of the GC
sources are IPs. To examine this idea, we include CVs in our
calculations, and set 5\% of them are IPs (Kube et al.
\cite{Kube}).

\subsubsection{ LMXBs}

This type of systems contain NSs or black holes (BHs) accreting
from low-mass companions that overfill their Roche lobes. Most of
LMXBs are transients, with the majority of their time spent in
quiescence with $L_{\rm X}<10^{34}\,\ergs$. The criterion
suggested by van Paradijs (\cite{van Paradijs}) is used to
determine whether the X-ray source is persistent or a soft X-ray
transient (SXT). BT04 took a semi-empirical approach for the X-ray
luminosities of LMXB transients in quiescence. We have instead
adopted the theoretical model of Menou et al. (\cite{Menou}) for
quiescent SXTs, in which the inner part of the accretion flow is
advection-dominated (ADAF, see Narayan, Mahadevan \& Quataert
\cite{Narayan98} for a review), surrounded by an outer thin disk.
For BH SXTs, the quiescent luminosities are
\begin{equation}
L_{\rm X,q}=g_{\rm BH}/(1+g_{\rm
BH})\times10^{-5}\times\eta\dot{M}c^2,
\end{equation}
where $g_{\rm BH}\equiv\dot{M}_{\rm ADAF}/\dot{M}_{\rm acc}\simeq
0.5$, $\dot{M}_{\rm ADAF}$ is the rate at which mass is accreted
via the ADAF, and $\dot{M}_{\rm acc}$ the rate at which mass is
accumulated in the outer thin disk. The accumulated mass will be
accreted in the outbursts.

Menou et al. (\cite{Menou}) included the propeller effect to
estimate the quiescent X-ray luminosities of NS transients
\begin{equation}
L_{\rm X, q}=g_{\rm NS}/(1+g_{\rm NS})\times f_{\rm
acc}\times\eta\dot{M}c^2,
\end{equation}
where $g_{\rm NS}\simeq 0.2$ and $f_{\rm acc}$ is the fraction of
the total mass through ADAF that reaches the NS surface.

Both normal star and white dwarf (WD) donors are considered in our
simulations. Apart from core collapse SNe, BT04 suggested the
formation of NSs via accretion-induced collapse (AIC) of massive
WDs. The EPS code in our simulation also includes this
possibility: steady transfer of He-rich or C-rich material onto an
ONe WD can lead to an AIC leaving an NS remnant, when the WD
masses exceed the Chandrasekhar mass limit.

\subsubsection{ Rotation-powered pulsars}

In the ejector stage, the spin-down of NSs is governed by the
canonical rotation-powered pulsar mechanism. The rotation power is
spent to generate the magnetic dipole waves and accelerate
particles. Young pulsars are X-ray emitters. There appears to be a
strong correlation between the rate of rotational energy loss
$\dot{E}$ and their X-ray luminosities $L_{\rm X}$ (Seward \& Wang
\cite{Seward}; Becker \& Trumper \cite{Becker}; Saito
\cite{Saito}). The comprehensive investigation by Possenti et al.
(\cite{Possenti}) suggested that the X-ray luminosities in $2-10$
keV band depend on $\dot{E}$ with the following relation
\begin{equation}
L_{\rm X}=10^{-15.3}(\dot{E}/\ergs)^{1.34}\,\ergs.
\end{equation}
Since  $\dot{E}$ decreases with time, rotation-powered pulsars can
shine in X-rays with luminosities $\sim {10^{31}-10^{35}}\ergs$ at
ages of $\sim{10^{3}-10^{4}}$ yr.  The X-ray flux contains both
pulsed non-thermal X-ray emission from the NS magnetosphere and
unpulsed component from the synchrotron nebulae powered by the
relativistic particles and magnetic fields ejected by the NS.
Contribution from the NS cooling is most likely in the energy band
$\leq 2$ keV and could be highly attenuated by the Galactic
interstellar absorption.

\subsubsection{ Massive stars with strong winds}

Massive stars with strong winds, like Wolf-Rayet stars, are long
known to be bright X-ray sources (Pallavicini et al.
\cite{Pallavicini}). The X-ray emission is generally thought to be
generated in the lower layers of the winds, where the plasma is
heated in shocks arising from small-scale wind structures that
grow out from line-driven wind instabilities (Feldmeier et al.
\cite{Feldmeier}; Dessart \& Owocki \cite{Dessart}). Chelbowski \&
Garmany (\cite{Chelbowski}, hereafter CG91) found that the X-ray
luminosities of both Wolf-Rayet stars and O type stars can be
fitted by the following relation,
\begin{equation}
   \log{L_{\rm X}}\,(\ergs)=34.32+0.34\log{(\dot{M}_{\rm w}v_{\infty})},
\end{equation}
where $\dot{M}_{\rm w}$ is the wind mass loss rate in units of
$\dot{M}_{\odot}\,{\rm yr}^{-1}$ and $v_{\infty}$ is the terminal
velocity of the mass outflows in units of kms$^{-1}$.

Massive binaries can add considerable X-rays due to collision of
the winds from the two stars under specific conditions. First,
this X-ray enhancement is significant only when the binary
separation $a$ lies in the following range (CG91),
\begin{equation}
   1.1(R_{1}+R_{2})\leq a \leq 5.5(R_{1}+R_{2}),
\end{equation}
where $R$ is the stellar radius, and the subscripts 1 and 2 denote
the primary and secondary stars, respectively. Second, Prilutskii
\& Usov (\cite{Prilutskii}) showed that the X-ray emission due to
wind collision in binary systems is inefficient unless
$\dot{M}_{2}v_{2}/(\dot{M}_{1}v_{1}+\dot{M}_2v_{2})\ge 0.4$. If
the above requirements are satisfied, we use the empirical
relation of Portegies Zwart et al. (\cite{Portegies02a}) to
calculate the X-ray luminosities for colliding winds,
\begin{eqnarray}
   L_{\rm X}&=&1.3\times10^{34}(\frac{\dot{M}_1}{10^{-5}M_{\odot}
   \mathrm{yr}^{-1}})^{0.4} \nonumber\\
   & &\times(\frac{v_1}{10^3\,
   \mathrm{kms}^{-1}})^{-0.65}(\frac{a}{R_{\odot}})^{-0.2}\ergs.
\end{eqnarray}

\subsection{X-ray Band and Luminosities}
Wang et al. (\cite{Wang}) survey revealed $\sim1,000$ discrete
sources, most of which were detected in the relative hard energy
range of 2-10\,keV. Less than 20 of these sources are previously
known (bright X-ray binaries).  They suggested that half of these
hard X-ray sources could be background active galactic nuclei
(AGN). Considering the steep density profile in the Galactic disk,
at least 100 sources with luminosities $10^{33}$\,-\,$10^{35}
\ergs$ (Pfahl et al. \cite{Pfahl}) are in the GCR.  Among the 2357
point sources presented by Muno et al. (\cite{Muno03}), 281
sources were detected below 1.5 keV and are believed to be mainly
in the foreground of the GC, and $\sim$100 sources are suggested
to be background AGN.  The surface density of the remaining
sources, and their absorption column toward the GC, in agreement
with those in infrared surveys toward Sgr A$^{*}$, demonstrate
that these X-ray sources trace the general stellar population at
the GCR (Muno et al. \cite{Muno03}).  A tiny fraction of the
remaining sources would be contaminated by the forground objects
because of the steep density profile of the Galactic disk. In
fact, the ratio of the forground and the GCR sources
$(8.5\mathrm{\,kpc}\times1)/(400\mathrm{\,pc}\times1000)\approx0.02$
(see \S 2.4 for the GCR stellar density prescription), is
negligible.  By using the typical spectral model obtained from
those relatively bright sources in this sample, Muno et al.
(\cite{Muno03}) inferred the 2.0-8.0 keV luminosity range of
$10^{30}\sim 10^{33}$erg s$^{-1}$ for the GC sources.  The
luminosities estimated for some of the models described in \S 2.1
are bolometric. They have to be converted into the 2-10 keV
luminosities  to be compared with observations.

WNSs often exhibit non-thermal spectra that can be described with
a $\Gamma\sim0$ power law below 10 keV (e.g. Campana et al.
\cite{Campana}), we thus adopt their 2-10 keV luminosities to be
half the calculated bolometric luminosities. Magnetized CVs,
especially IPs, usually exhibit hard thermal spectra ($kT\geq$10
keV) (e.g. Ezuka \& Ishida \cite{Ezuka}). We also take their
bolometric correction factor to be 0.5. The theoretical model of
Menou et al. (\cite{Menou}) of quiescent X-ray transients is
established for comparison with observational luminosities in
0.5-10 keV. So we multiply them by a factor of 0.5 to estimate the
luminosities in 2-10 keV.  The X-ray luminosities determined by
Eq.(7) for rotation-powered pulsars are appropriately in 2-10 keV
, therefore no correction is needed. CG91 obtained the empirical
relation (Eq.8) for massive stars with strong winds from the data
of {\em Einstein} observatory (in 0.5-3.5 keV).  For comparison,
Portegies Zwart et al.(\cite{Portegies02a}) calculated the
luminosities in the 0.5-3.5 keV band (Eq.10) based on the archival
{\em Chandra} X-ray observations. Taking account of their soft
spectra ($kT\sim1-3 \mbox{ keV}$, Portegies Zwart et al.
\cite{Portegies02a}), we choose the correction factor to be 0.1
for their 2-10 keV luminosities.

The X-ray emission is generally assumed to be isotropic. However,
for disk accreting sources, the geometric effect may affect the
apparent luminosity distributions (Zhang, \cite{Zhang}).
Quantitatively, the intrinsic luminosity is reduced by a factor of
$\cos \theta(1+2\cos \theta)/3$, where $\theta$ is the inclination
angle of the accretion disk; the factor of $\cos \theta$ is due to
the area-projection effect, and the factor of $(1+2\cos\theta)/3$
is due to the limb-darkening effect (Netzer, \cite{Netzer}).  The
probability of seeing an accretion disk at an inclination angle
$\theta$ is proportional to $\sin\theta$ if accretion disks are
assumed to be oriented randomly in the sky. Therefore, the
convolution between $f(x)=(1-x^{2})^{1/2}[1+2(1-x^{2})^{1/2}]$
(where $x=\sin\theta$ is uniformly distributed between 0 and 1)
and a given intrinsic luminosity distribution produces the
observed distribution (see e.g. Figures 2, 5 and 6).  Thus, we
take this inclination effect into account for Roche lobe overflow
(RLOF) X-ray binaries (i.e. IP and LMXBs).

\subsection{Star Formation and IMF in GCR}
As noted by Muno et al. (\cite{Muno04a}), it is still a matter of
debate as to whether the star formation is continuous or episodic,
and whether it occurs only in localized regions or is relatively
uniform throughout the GC. By modeling the evolution of the
population of luminous infrared stars, Figer et al.
(\cite{Figer04}) suggested that the stat formation is probably
continuous over the last $\sim10$ Gyr, as we did in this paper.
The Kroupa et al. (\cite{Kroupa}) IMF adopted in this work has a
slope of $-2.7$ when $M>1.0\,M_{\odot}$, derived from the stellar
distribution towards both Galactic poles as well as the
distribution of stars within 5.2 pc from the Sun. However, the
molecular clouds in the GCR have higher densities and temperatures
than those in the disk due to the tidal forces. This environment
might favor the formation of massive stars (Morris \cite{Morris}).
The {\em Hubble Space Telescope} observation of Arches Cluster
(which is within Wang et al. field) suggested an fairly flat IMF
with a slope of $\sim-1.7$\,-\,$-1.9$ (Figer et al.
\cite{Figer99}, Figer \cite{Figer05}; see also Stolte et al.
\cite{Stolte}). Portegies Zwart et al. (\cite{Portegies02b}),
however,  showed that the observed characteristics (unusually flat
mass function and overabundance of massive stars) of the Arches
cluster are consistent with a perfectly normal IMF. The observed
anomalies are then caused by a combination of observational
selection effects and the dynamical evolution of the cluster.
Considering lack of the knowledge of the star formation history in
GCR, our assumptions of star formation and IMF in the GCR are not
incompatible with recent observational and theoretical works.

\subsection{Expected Numbers}

The expected numbers for each class of X-ray sources in the GCR
are related with the the ratio of the star population in the GCR
and in the total disk. The size of Wang et al. survey field is
$2^{\circ}\times 0.8^{\circ}$. Supposing the space density of
stars in the Galactic disk to be
$n(R,z)\propto$exp$(-R/R_{0})$exp$(-|z|/z_{0})$ and integrating
$n(R,z)$ over 1.6 deg$^{2}$ through the GCR, Pfahl et al.
(\cite{Pfahl}) found that the field contains $\leq$1\% of the
total Galactic disk population.  Alternatively, we can make an
order of magnitude estimate on this ratio. Wang et al. survey
field encompasses a physical area of 300 pc by 120 pc at a
distance of 8.5 kpc.  Assume that all of the point sources are
within the nuclear bulge (which is about 300 pc across in the
radial direction; Mezger et al., \cite{Mezger}) and the average
stellar density is 100 M$_{\odot}$pc$^{-3}$ (which is 1000 times
that in the local neighborhood, Binney \& Merrifield
\cite{Binney}), the Wang et al. field contains a stellar mass of
300$\times$120$\times$300$\times$100=10$^{9}$M$_{\odot}$.  Thus,
this field contains 1\% of the total Galactic stellar mass, which
is $\sim10^{11}$M$_{\odot}$.  With EPS calculations, we try to
search various types of X-ray sources (described in Sec.2) that
can be detected in Wang et al. (\cite{Wang}) and Muno et al.
(\cite{Muno03}) surveys. After we get the total number $N_{\rm
Gal}$ of a specific type of sources in the Galactic disk, then
$\sim N_{\rm Gal}/100$ and $\sim{N_{\rm Gal}/100/4}$ sources of
this population are expected to be found in Wang et al. and Muno
et al. surveys, respectively, since the latter field contains
stars 4 times less than the former (BT04). Note that sources found
in these two surveys have different luminosity ranges.

\section{Results}

We adopt a variety of models (see Table 1), each with different
assumptions for the parameters that govern the evolutions in the
calculations.  Tables 2 and 3 summarize the calculated numbers of
various classes of X-ray sources contributed to Wang et al. and
Muno et al. surveys, respectively.  The expected numbers after
correcting for the inclination effect are also listed.

The luminosity distributions of WNSs are shown in Figure 1. The
solid and dashed lines represent the calculated results with and
without the propeller effect considered, respectively. The latter
distribution implies that there are about 12 and 16 WNSs (for
Models A and D respectively) in Wang et al. field. PRP02 predicted
$\sim250$ WNSs in the solid angle of Wang et al field for their
standard kick model, and $\sim 12-100$ of them could be detected
in the survey. This is compatible with our result (we get 25 WNSs
in Model F, in which the SN kick model is same as the standard
model of PRP02).  The figure also reveals that with decreasing
wind velocities the WNS luminosities without considering the
propeller effect increase, as expected with the Bondi-Hoyle
mechanism.  {\em However, when we take account of the propeller
effect, the expected number of WNSs that can be detected in Wang
et al. survey is less than one, indicating that mass accretion may
be highly inefficient during most part of the WNS stage.}

Figure 2 shows the luminosity distributions of various classes of
RLOF X-ray binaries expected in Muno et al. field. The thin solid
lines, dashed lines and thick solid lines are for IPs, NS LMXBs
and BH LMXBs respectively. In this figure we can see that NS LMXBs
in quiescence, generally brighter than BH LMXBs, account for a
considerable fraction of the sources in Wang et al. survey (30 out
of $\sim 100$ for model F, see Table 2);  NS LMXBs with WD donor
stars contribure a significant fraction of the sources in Muno et
al. field. Most of the NS-WD LMXBs are ultracompact, with orbital
periods less than 2 hr. Their numbers (see Table 3) are roughly in
line with that (291) in BT04.  Our simulations predict hundreds of
IPs in the GCR, fairly less than Muno et al. (\cite{Muno04a})'s
estimate ($\sim1000$).

The CE efficiency parameter $\alpha_{CE}$ plays an important role
in the evolution of close binaries. To see the dependence of the
results on $\alpha_{CE}$, we can compare the results of Model A
($\alpha_{CE}=1.0$) with those of Model B ($\alpha_{CE}=0.5$) in
Table 3. The parameter $\alpha_{CE}$ has two contrary effects on
the evolution of close binaries.  On one hand, high value of
$\alpha_{CE}$ results in relatively wide binaries after the CE
phase, so that the following RLOF mass transfer may be hard to
take place. On the other hand, for the ultracompact X-ray
binaries, a high value of $\alpha$ can prevent coalescence during
the previous CE phase, significantly increasing the formation rate
of these systems. The overall influence is that larger $\alpha$
leads to more LMXBs, especially LMXBs with a WD donor, because
these binaries have experienced CE evolution at least twice.

Both IP and LMXB numbers increase significantly when the tidal
spin-orbit coupling is considered in Models E and G.  When the
primary star ascends to the giant branch, its expansion causes a
reduction of the orbital separation through the spin-orbit
interaction, while wind mass loss from the evolved donor star
causes the orbit to widen.  The competition between the wind loss
and the tidal interactions will determines the subsequent orbital
evolution (Tauris \& Savonije \cite{Tauris1}; Tauris
\cite{Tauris2}).  As a typical example of CV formation in Model E
(TIDE ON) let us consider binary stars of masses 2.9 and
0.4\,M$_{\odot}$ in an orbit with initial separation of
484\,R$_{\odot}$. About 0.5\,Gyr later, the more massive star
evolves to the AGB stage when the separation is 488\,R$_{\odot}$.
In the following 3\,Myr evolution , the tidal spin-orbit
interaction dominates the orbital evolution and the binary
separation decreases by 30\% to 355\,R$_{\odot}$ when the primary
star begins to fill its Roche lobe. A CE forms and a CO white
dwarf separated from the 0.4\,M$_{\odot}$ main sequence (MS) star
by 2\,R$_{\odot}$ emerges.  When the separation has fallen to
1.1\,R$_{\odot}$ (due to tidal interaction) 70\,Myr later, the
secondary star fills its Roche lobe and the stage of CV evolution
begins.  In contrast, the same system in Model A (TIDE OFF) has an
ever increasing orbital separation (from 484 to 500\,R$_{\odot}$,
owing to the wind mass loss effect) before the primary star fills
its Roche lobe.  The CE evolution leads to a CO WD-MS star binary
with a separation of 3\,R$_{\odot}$.  The secondary star cannot
fill its Roche lobe within 12\,Gyr because the timescale for
either its nuclear expansion or the orbital shrinkage due to
gravitational radiation is too long.

In Figure 3 we show the X-ray luminosity distributions of
rotation-powered pulsars. There are totally $\sim$28 and $\sim$9
pulsars predicted in Muno et al. and Wang et al. surveys
respectively.  The numbers of rotation-powered pulsars rely only
on the core-collapse SN rate and their X-ray lifetime, both are
independent of the parameters in Table 1.

Figure 4 shows the X-ray luminosity distributions of massive stars
with strong winds.  Only a small fraction of them have colliding
winds and thus are more luminous.  Their luminosities increases
with decreasing wind velocities. The reason is that the X-ray
luminosity from colliding winds is proportional to the product of
the emissivity per unit volume, $n^{2}\Lambda$, both the density
$n$ and emission rate $\Lambda$ are anticorrelated with the wind
velocity $v_{W}$. We obtain $\sim 70$ massive stars in Muno et al.
survey. The number hardly changes with the model parameters.

Figures 5 and 6 show the X-ray luminosity distributions of the
total X-ray sources for different models.  All models (especially
models E and G with TIDE ON) seems to produce X-ray sources with
reasonable numbers in Muno et al. field, but less than half the
objects in Wang et al. field are accounted for.  In Muno et al.
(\cite{Muno03}) the luminosity distribution peaks around
$2\times10^{31} \ergs$, nearly one order of magnitude higher than
in our work.  The reasons for this discrepancy can be addressed
briefly as follows.  First the calculated mass transfer rates are
long-term, averaged ones.  They may not be directly compared with
observed, instantaneous X-ray luminosities.  Second, observational
selection effects may favor relatively more luminous X-ray sources
to be identified.  Third, some potentially important effects (such
as the effect of X-ray irradiation on the secondary) have not been
included in our binary calculations, which could considerably
alter the evolution of these systems and increase the mass
accretion rates (see also Podsiadlowski, Rappaport \& Pfahl
\cite{Podsiadlowski}).

We also calculate the numbers of bright X-ray sources in GCR. The
outburst correction factor is chosen to be 0.1 and 1 for the
short- and long-period LMXB transient systems, respectively (see
BT04 for details).  We assume the bolometric correction factor is
0.1 and the duty cycle is 1\%.  For Model A, 23 X-ray sources
brighter than 10$^{36}\ergs$ are found in Muno et al. field,
including 16 transient LMXBs in outburst and 7 persistent LMXBs.
Only a few bright transients are detected in Muno et al. survey.
For the Wang et al. field, we predict 92 bright point sources,
whereas $\leq$20 bright sources were detected in Wang et al.
(\cite{Wang}).  Like BT04, our simulation seems to lead to an
overproduction of the bright systems. However, for model B, the
expected numbers of bright sources in Muno et al. and Wang et al.
surveys are reduced to 9 and 36, respectively.

We calculate the point source contribution to the X-ray emission
that has previously been ascribed to diffuse emission (Koyama et
al. \cite{Koyama}; Sidoli \& Mereghetti \cite{Sidoli}) in the GCR.
For various models, the X-ray point sources ($\leq10^{35} \ergs$)
produce a mean surface brightness of $1.3-2.6\times10^{-14}$erg
cm$^{-2}$ s$^{-1}$ arcmin$^{-2}$.  This is 3\%-9\% of that of the
diffuse emission from the inner regions of the Galaxy derived by
Sidoli \& Mereghetti (\cite{Sidoli}) and by Koyama et al.
(\cite{Koyama}). Our result is compatible with the observational
result (10\%) of Muno et al. (\cite{Muno03}).

\section{ Discussion and conclusions}

By use of the EPS method, we have investigated the nature of the
faint X-ray sources in the GCR. Our simulated objects include
WNSs, magnetic CVs, LMXBs, rotation-powered pulsars and massive
stars with strong winds. The main results are summarized in Tables
2 and 3, with $\sim$10-30 and $\sim$1000-2000 sources expected to
be detected in Wang et al. and Muno et al. surveys, respectively.
For Wang et al. survey, we find that a considerable fraction of
the discrete sources may be rotation-powered pulsars and NS LMXB
transients in quiescence, while WNSs proposed by PRP02 have
negligible contribution due to the propeller effect (Note that we
have not considered Be/X-ray binaries in which the NS evolution
may be quite different from investigated here.). Recent near
infrared imaging of the X-ray sources in the GCR shows that the
colour distribution of the identified candidate counterparts of
some of the X-ray sources is redder than expected for WNS systems,
but consistent with later-type stars (Bandyopadhyay et al.
\cite{Bandyopadhyay}). For Muno et al. field, IPs present a minor
contribution, and the majority of the X-ray sources seem to be NS
LMXB transients with WD donors. The latter result is consistent
with BT04.

Our calculations also suggest that some of the point sources
detected by Muno et al. survey may be massive stars with strong
winds. Radio emission from these winds should be detectable at
centimeter wavelengths (Panagia \& Felli \cite{Panagia}; Wright \&
Barlow \cite{Wright}).  We propose that a systematic radio
observing campaign be undertaken to search for the stellar
counterparts in the GCR surveyed by Muno et al. (some efforts have
been made, e.g. by Lang et al. \cite{Lang} most recently).
Moreover, radio observations could reveal the existence of jets,
which are common in X-ray binaries.

Sakano et al. (\cite{Sakano}) recently reported the discovery in
the GCR of two unusual X-ray transients XMM J174457$-$2850.3 and
XMM J174544$-$2913.0 with flux variations in excess of a factor of
100 during roughly a year and peak X-ray luminosities of
$\sim5\times10^{34}\,\ergs$ and with peculiar spectral features.
Since no known classes of sources can well explain all their
characteristics, these authors argued that these two sources may
represent a new type of sources with different properties from
those we have known. This also implies that what is happening in
the GCR is much more complicated than we have already learnt.

\begin{acknowledgements}
We would like to thank Jarrod R. Hurley for kindly providing us
his SSE and BSE code and valuable conversations. We thank Z.-R.
Wang, Y. Chen, Z.-W. Han and M. Li for useful discussions.  We are
grateful to the anonymous referee for constructive suggestions for
improving the manuscript. This work was supported by the National
Natural Science Foundation of China under grant 10025314 and the
Ministry of Science and Technology of China under grant NKBRSF
G19990754.
\end{acknowledgements}

\clearpage

  \begin{figure}
   \centering
   \includegraphics[width=15cm]{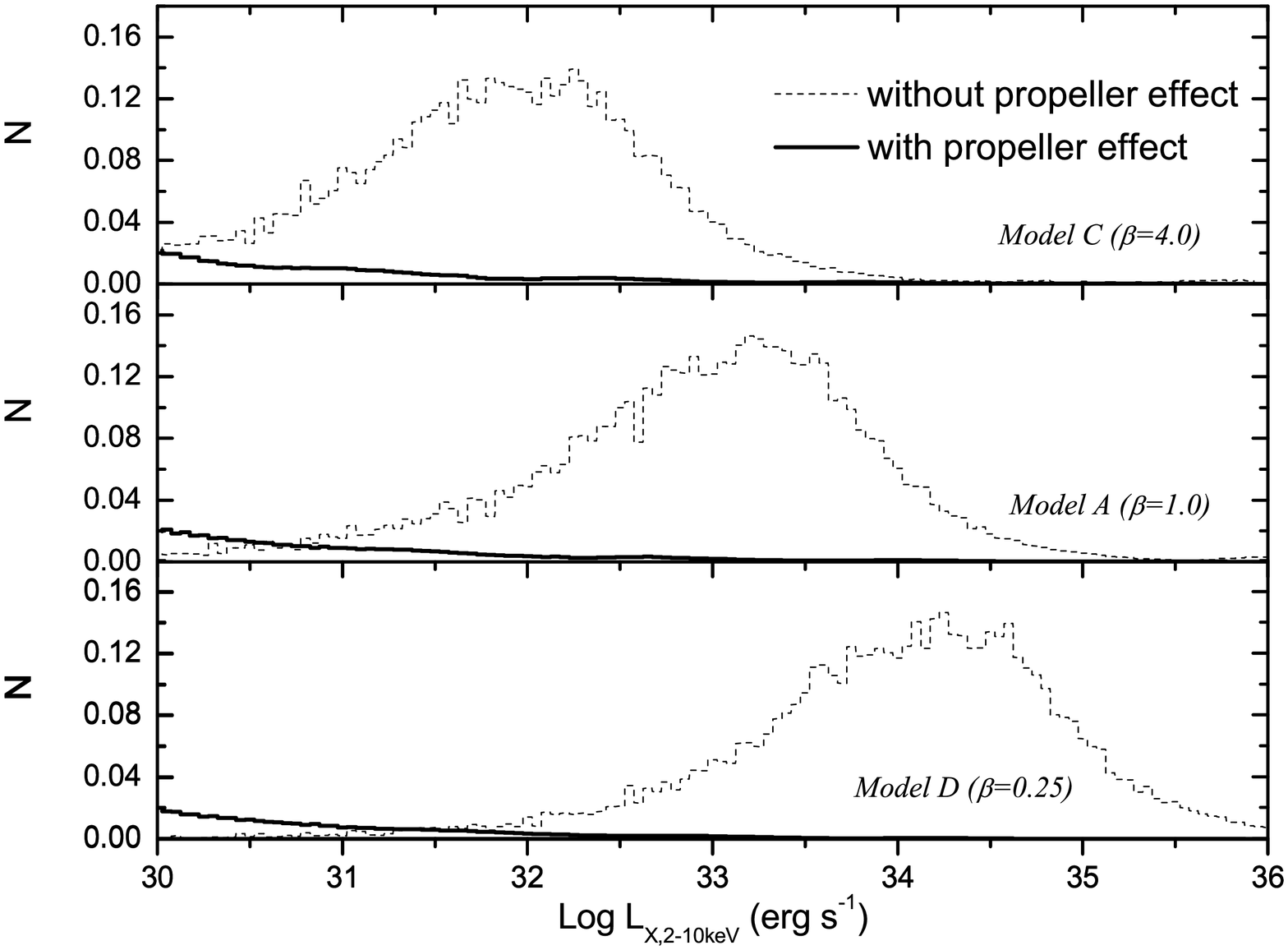}
      \caption{The X-ray luminosity distributions for WNSs with and without
      considering the propeller
      effect.  Three Models C, A and D with decreasing wind
      velocity are shown from top to bottom.
              }
   \end{figure}

\clearpage

  \begin{figure}
   \centering
   \includegraphics[width=15cm]{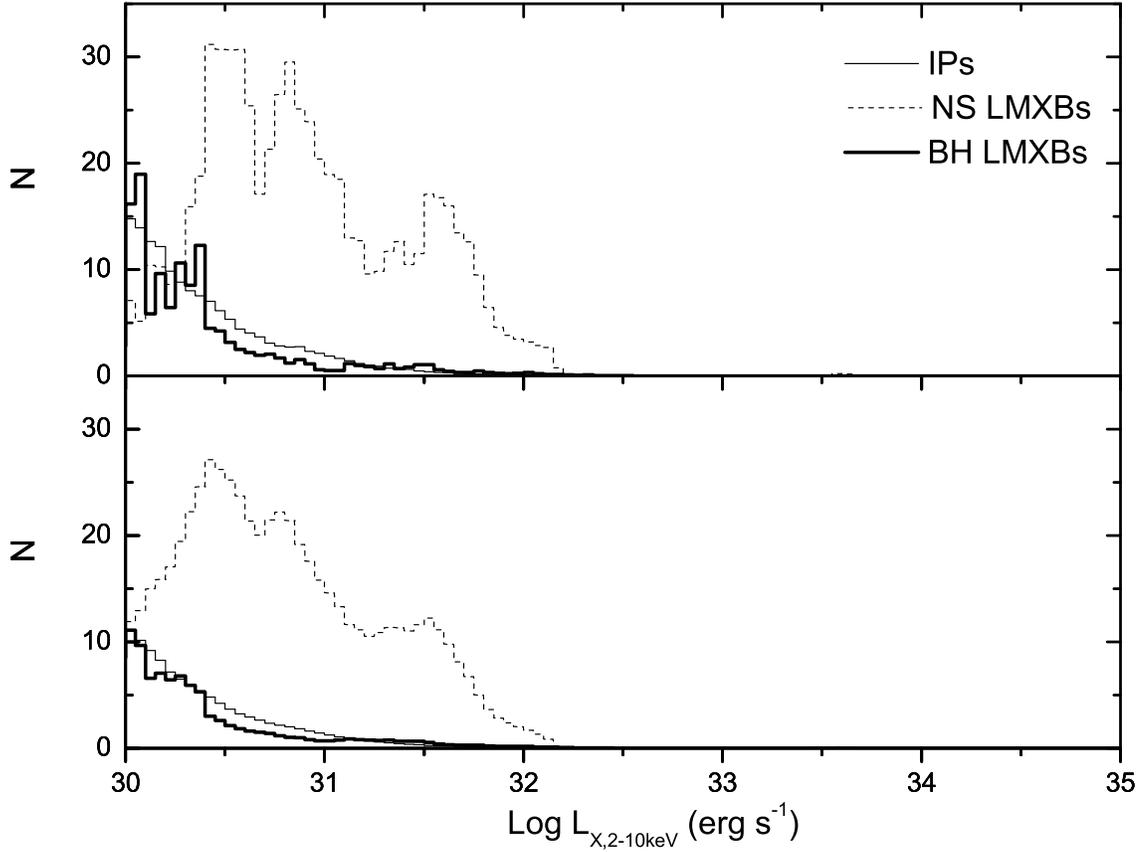}
      \caption{The X-ray luminosity distributions for Roche-lobe
      overflow systems including IPs (thin solid lines), NS LMXBs (dashed lines) and
      BH LMXBs (thick solid lines).  The figure is plotted only for Model A since the main
features do not change much as the parameters change.  Upper
panel: intrinsic
      distributions.  Lower panel: distributions after correcting for accretion disk inclination
      effect.
              }
   \end{figure}

\clearpage

  \begin{figure}
   \centering
   \includegraphics[width=15cm]{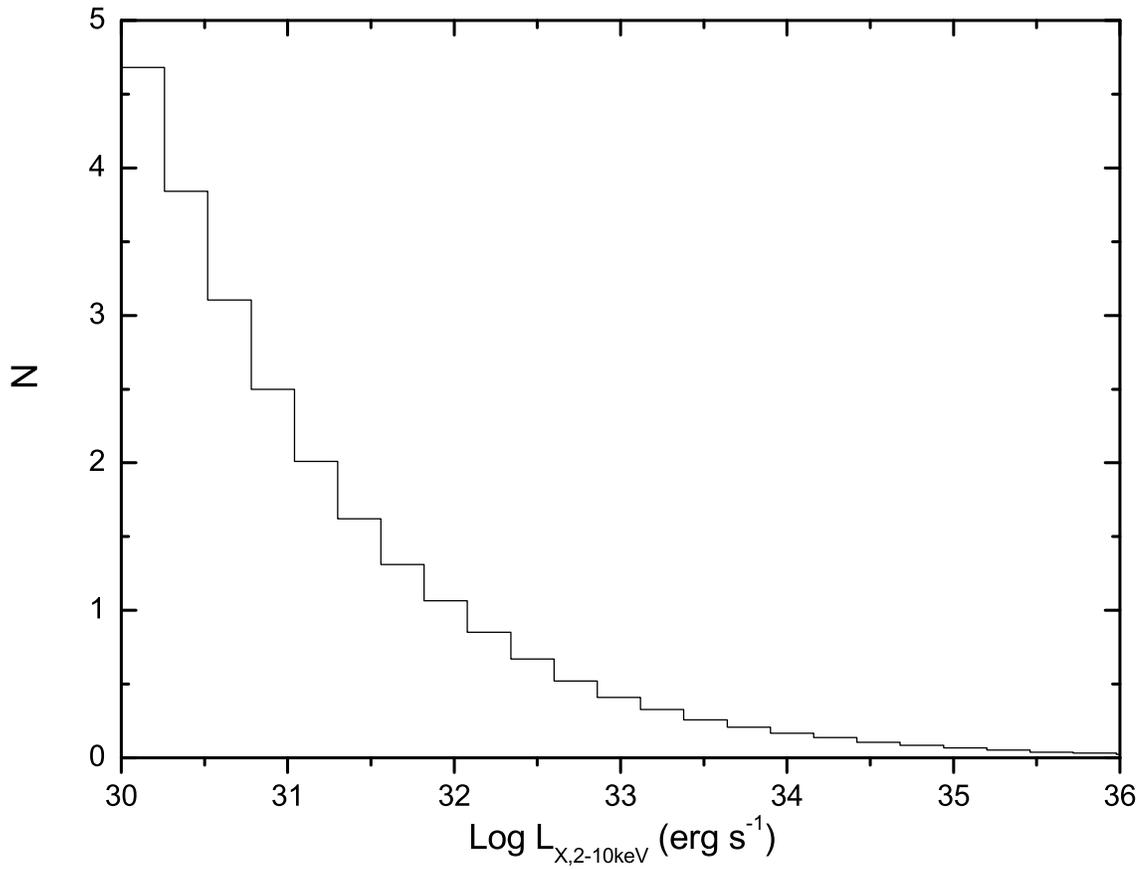}
      \caption{The X-ray luminosity distributions for rotation-powered
pulsars.  This figure is plot for Model A. Distributions for other
models have little changes.
                    }
   \end{figure}

\clearpage

  \begin{figure}
   \centering
   \includegraphics[width=15cm]{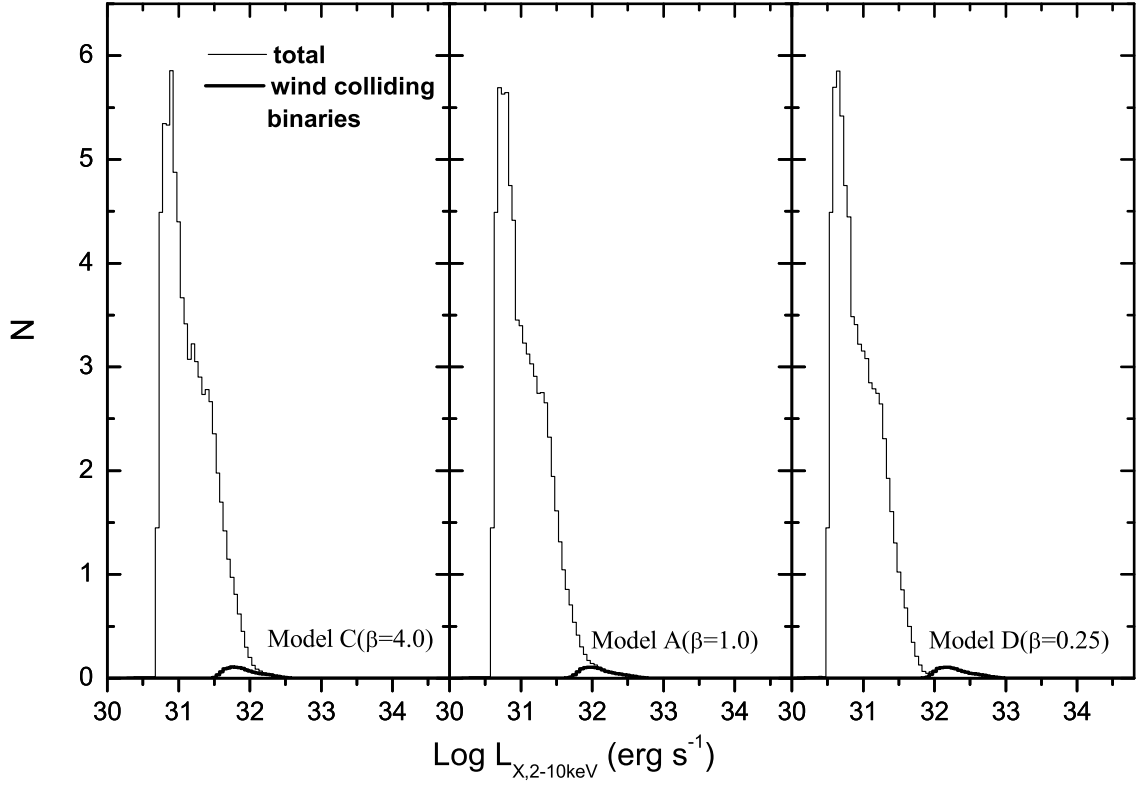}
      \caption{The X-ray luminosity distributions for massive stars
      with strong winds (dashed lines).  Distributions for wind-colliding binaries are highlighted with thick
      lines. Three Models C, A and D with decreasing wind
      velocity are shown from left to right.
                    }
   \end{figure}

\clearpage

  \begin{figure}
   \centering
   \includegraphics[width=15cm]{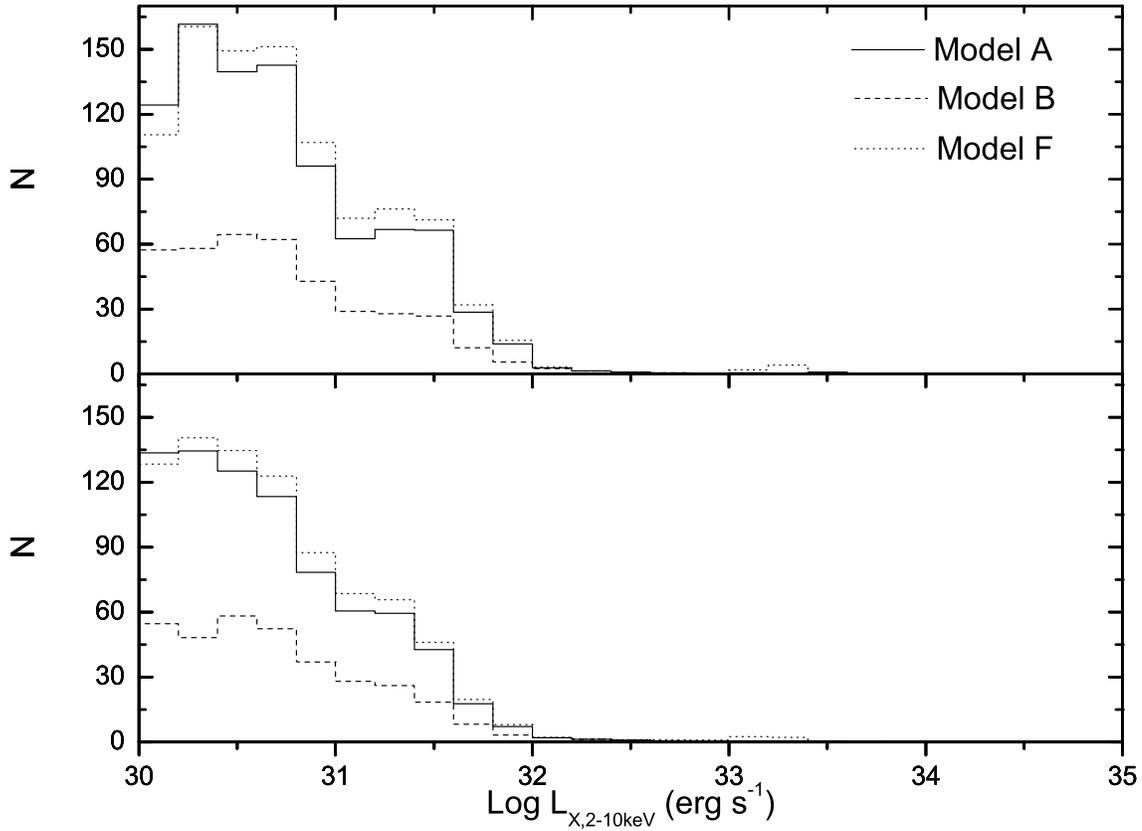}
      \caption{The total number of
      X-ray sources as a function of the X-ray luminosity.
      Three models, A (solid lines), B (dashed lines) and F (dotted lines),
      in which no tidal effect considered are shown.  Upper
      panel: intrinsic
      distributions.  Lower panel: distributions after correcting for the accretion disk inclination
      effect.
                    }
   \end{figure}

\clearpage

  \begin{figure}
   \centering
   \includegraphics[width=15cm]{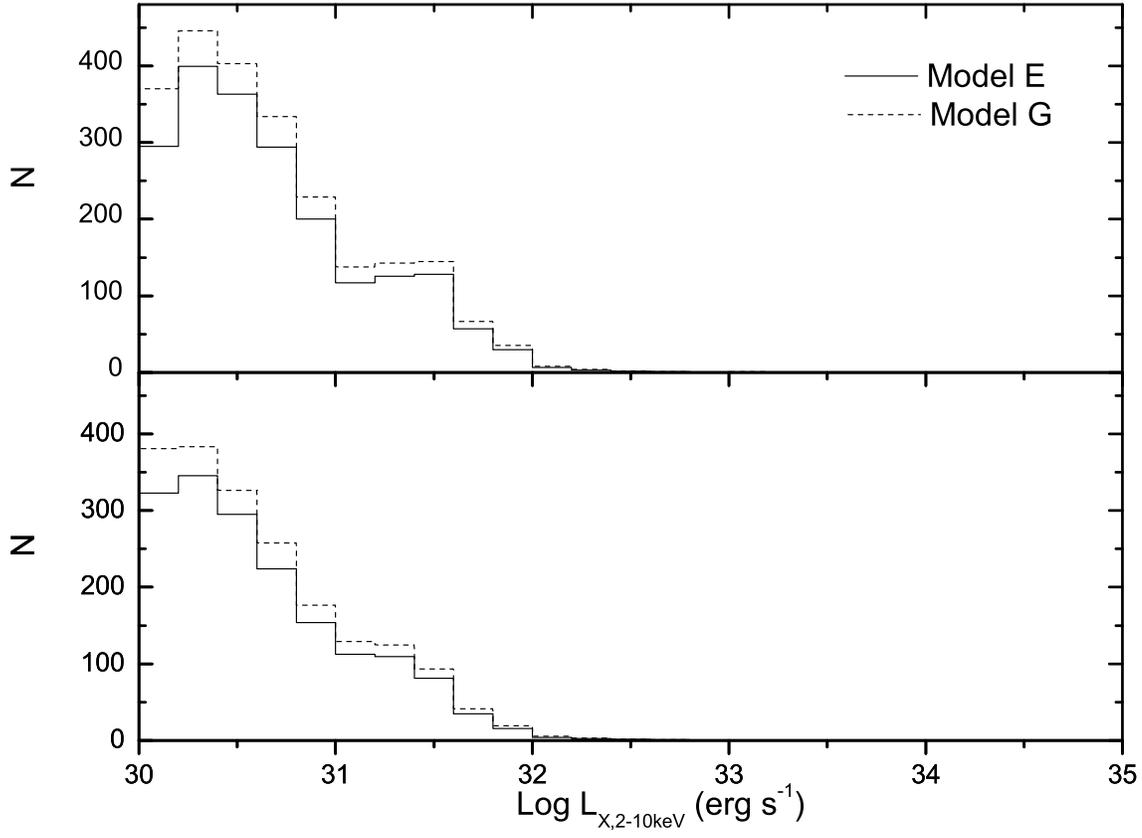}
      \caption{Same as Fig.5.  Two models, E (solid lines) and G (dashed lines),
      in which tidal effect was considered are shown here.  Upper
      panel: intrinsic
      distributions.  Lower panel: distributions after correcting for the accretion disk inclination
      effect.
                    }
   \end{figure}

\clearpage

\begin{table}
\caption{Model parameters.} \centering
\begin{tabular}{c|cccc}\hline\hline
   Model & $\alpha_{CE}$ & $\beta$ & $\sigma$ & tides\\ \hline
      A & 1.0 & 1.0 & 265 & OFF\\
      B & 0.5 & 1.0 & 265 & OFF\\
      C & 1.0 & 4.0 & 265 & OFF\\
      D & 1.0 & 0.25 & 265 & OFF\\
      E & 1.0 & 1.0 & 265 & ON\\
      F & 1.0 & 1.0 & 190 & OFF\\
      G & 1.0 & 1.0 & 190 & ON\\ \hline

\end{tabular}
\end{table}

\begin{table}
\caption{The expected numbers of X-ray sources in Wang et al.
survey with $L_{\rm X,2-10keV}=10^{33}- 10^{35}\,\ergs$.  In
parentheses we list the expected numbers of X-ray binaries with WD
donors.  The expected numbers after correcting for the
inclination effect are listed in the last column.}             
\centering
\begin{tabular}{c|ccccc|c|c}\hline\hline
              Model & WNSs & NS LMXBs & BH LMXBs & IPs & Pulsars & TOTAL & TOTAL-inc \\ \hline
    A & $<1$ & 3(0) & $<1$(0) & 1 & 9 & 13 & 13\\
    B & $<1$ & $<1$(0) & $<1$(0) & 1 & 9 & 9 & 9\\
    C & $<1$ & 4(0) & $<1$(0) & 1 & 9 & 14 & 14\\
    D & $<1$ & $<1$(0) & $<1$(0) & 2 & 9 & 11 & 11\\
    E & $<1$ & 2(0) & 1(0) & 1 & 9 & 13 & 12\\
    F & $<1$ & 23(0) & $<1$(0) & 1 & 9 & 33 & 30\\
    G & $<1$ & 5(0) & 1(0) & 1 & 9 & 16 & 14\\
    \hline

\end{tabular}
\end{table}

\begin{table}
\caption{The expected numbers of X-ray sources in Muno et al.
survey with. $L_{\rm X,2-10keV}=10^{30}-10^{33}\,\ergs$.  In
parentheses we list the expected numbers of X-ray binaries with WD
donors.  The expected numbers after correcting for the
inclination effect are listed in the last column.}           
\centering
\begin{tabular}{c|ccccc|c|c} \hline\hline
         Model & IPs & NS LMXBs & BH LMXBs &
           Pulsars & Massive Stars & TOTAL & TOTAL-inc\\ \hline
  A & 165 & 648(646) & 132(0) & 28 & 72 & 1045 & 912\\
  B & 169 & 195(194)  & 7(0) & 28 & 71 & 470 & 405\\
  C & 163 & 642(641) & 130(0) & 28 & 72 & 1035 & 905\\
  D & 176 & 935(934) & 158(0) & 28 & 72 & 1369 & 1202\\
  E & 706 & 1142(1141) & 270(0) & 28 & 72 & 2218 & 1963\\
  F & 165 & 694(693) & 128(0) & 28 & 72 & 1087 & 961\\
  G & 706 & 1242(1240) & 497(0) & 28 & 71 & 2544 & 2248\\
  \hline
\end{tabular}
\end{table}

\end{document}